\documentclass{moriond}

\usepackage{amsmath}
\usepackage{subfig}
\usepackage{lineno}

\newcommand{\Photo}{} 
\newcommand{\ALL}{A_{LL}}
\newcommand{\ALLJET}{A_{LL}^{Jet}}
\newcommand{\ALLPEAK}{A_{LL}^{Peak}}
\newcommand{\ALLPI}{A_{LL}^{\pi^{0}}}
\newcommand{\ALLSIDE}{A_{LL}^{Side}}
\newcommand{\HGC}{\Delta{}G}
\newcommand{\HGD}{\Delta{}g}
\newcommand{\PI}{\pi^{0}}
\newcommand{\PT}{p_{T}}
\newcommand{\XT}{x_{T}}

\begin{document}

\vspace*{4cm}
\title{DOUBLE HELICITY ASYMMETRY IN $\PI$ PRODUCTION AT MIDRAPIDITY IN POLARIZED $p+p$ COLLISIONS AT $\sqrt{s}=510$ GeV}
\author{INSEOK YOON for the PHENIX COLLABORATION}
\address{The Research Institute of Basic Science, Seoul National University, 1, Gwanak-ro, Gwanak-gu, Seoul, 08826, Republic of Korea}

\maketitle

\abstract{
  PHENIX measurements are presented for the cross-section and double-helicity asymmetry ($\ALL$) of inclusive $\PI$ production ($\ALLPI$) at midrapidity from $p+p$ collisions at $\sqrt{s}=510$ GeV from data taken in 2012 and 2013 at the Relativistic Heavy Ion Collider (RHIC). The next-to-leading order (NLO) perturbative QCD (pQCD) calculation agrees excellently with the presented cross-section result. The $\ALLPI$ follows an increasingly positive asymmetry as functions of $\PT$ and $\sqrt{s}$ at the fixed $\XT$. The latest global analysis results, which support the positive spin contribution of gluon ($\HGC$), agrees excellently with the presented asymmetry result. The asymmetry result extends the experimental sensitivity to the previously unexplored $x$ region down to $x\sim0.01$ and provides additional constraints on $\HGC$.     
}

\section{Motivation}
Since the EMC experiment \cite{Ashman19891} showed that the spin contribution of quarks ($\Delta\Sigma$) to the proton spin is strikingly small, it has been revealed that understanding $\HGC$ is very important to understand the spin structure of the proton. Along with several measurements of polarized deep inelastic scattering (DIS) and polarized semi-inclusive DIS (SIDIS), RHIC polarized $p+p$ collisions and PHENIX measurements of $\ALL$ of inclusive $\PI$ ($\ALLPI$) at $\sqrt{s}=62.4$ GeV \cite{PhysRevD.79.012003} and $\sqrt{s}=200$ GeV \cite{PhysRevD.90.012007} and STAR measurements of $\ALL$ of inclusive jets ($\ALLJET$) at $\sqrt{s}=200$ GeV \cite{PhysRevLett.115.092002} constrained the helicity gluon distribution ($\HGD$) successfully. Resultingly, the QCD global analyses have observed positive $\HGC$ \cite{Nocera2014276,PhysRevLett.113.012001}. 

However, a large uncertainty remained in $\HGD$, especially in the small $x$ region, limits the understanding of $\HGC$. Thus expanding the experimental sensitivity to smaller $x$ region is very important. To access the smaller $x$ region, PHENIX measures $\ALLPI$ at a higher collision energy, $\sqrt{s}=510$ GeV \cite{PhysRevD.93.011501}. The $\ALLPI$ measurements at $\sqrt{s}=510$ GeV can access a smaller $x$ region $0.01<x$ while the $\ALLPI$ measurements at $\sqrt{s}=62.4$ GeV and at $\sqrt{s}=200$ GeV can access $x$ regions $0.06<x$ and $0.02<x$ respectively and the $\ALLJET$ measurement at $\sqrt{s}=200$ GeV can access an $x$ region $0.05<x$.

\section{Definition and Interpretation of $\ALL$}
\subsection{Factorization and $\PI$ Cross-Section}
The $\PI$ cross-section can be understood through factorization. The QCD factorization theorem allows to separate the cross-section into two parts: partonic reaction cross-section ($\hat{\sigma}$) which governs short-distance physics and is calculable via pQCD and long-distance functions such as parton distribution functions (PDF) and fragmentation functions (FF) which are uncalculable but universal.

\begin{figure}[htb]
  \centering
  \includegraphics[width=0.4\textwidth]{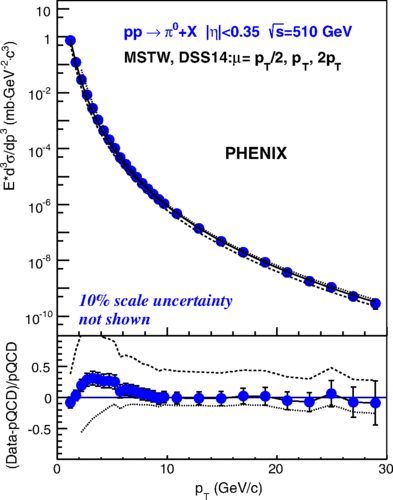}
  \caption{Midrapidity $\PI$ production cross-section at $\sqrt{s}=510$ GeV as function of $\PT$. NLO pQCD calculation with theory scale $\mu=\PT/2$ (dotted line), $\PT$ (solid line) and $2\PT$ (dashed line). (top) Relative difference between the data and theory. (bottom) \protect\cite{PhysRevD.93.011501}}
  \label{fig:pi0_crosssection}
\end{figure}

To check the validity of the factorization, midrapidity $\PI$ cross-section at $\sqrt{s}=510$ GeV is measured as Figure~\ref{fig:pi0_crosssection}. The experimental $\PI$ cross-section is compared to NLO pQCD calculations performed with MWTS 2008 PDFs \cite{Martin2009} and DSS14 FFs \cite{PhysRevD.91.014035}. That excellent agreement assures that factorization is valid and the our understanding of parton-to-hadron fragmentation becomes mature.

\subsection{Definition and Interpretation of $\ALL$}
The $\ALL$ of final state hadron $C$ in longitudinally polarized proton collisions, $p+p\to{}C+X$, can be defined in terms of differences in cross-section as 
\begin{equation} \label{eq:all_def}
  \ALL=\frac{d\Delta\sigma}{d\sigma}=\frac{d\sigma_{++}-d\sigma_{+-}}{d\sigma_{++}+d\sigma_{+-}}
\end{equation}
where $\sigma_{++~(+-)}$ stands for hadron production cross-section in same (opposite) proton helicity collisions.

The polarized cross-section, $\Delta\sigma$ can be factorized into:
\begin{equation} \label{eq:factorization}
  \Delta\sigma=\sum_{a,b,c}\Delta{}f_{a}(x_{a},\mu)\otimes\Delta{}f_{b}(x_{b},\mu)\otimes\Delta\hat{\sigma}_{ab}^{c}(x_{a}P_{a}, x_{b}P_{b}, P_{C}/z_{c}, \mu, \mu^{\prime})\otimes{}D_{c}^{C}(z_{c}, \mu^{\prime})
\end{equation}
where $\Delta{}f=f_{+}(x,\mu)-f_{-}(x,\mu)$ is a helicity PDF describing the difference in density of partons being aligned (+) and anti-aligned (-) with the proton's helicity at a certain Bjorken $x$. $\hat{\sigma}_{ab}^{c}$ is a partonic cross-section for the process $a+b\to{}c$. $D_{c}^{C}(z_{c}, \mu^{\prime})$ is a FF of a parton c into a final state hadron C at a fractional energy $z_{c}$. The unpolarized cross-section can be factorized in a similar way.

$\ALLPI$ is a good probe to access $\HGD$ because not only the $\PI$ cross-section is well understood as shown in Figure~\ref{fig:pi0_crosssection} but also $\PI$ at mid-rapidity is predominantly created in gluon-gluon and quark-gluon scattering. 

\section{Introduction to PHENIX Spin Runs and Configuration}
RHIC is the world's only one polarized proton collider and the unique facility to explore the spin structure of the proton. The direction of the proton spin can be controlled at the bunch level. Proton bunches with every combination of the spin directions collide within 8 bunch crossings$\sim$848 ns. It helps to suppress the occurrence of any systematic uncertainty due to variation of trigger efficiency or detector acceptance. 

Since 2003, PHENIX has taken several spin runs including not only longitudinal but also transverse runs. Longitudinal spin data at $\sqrt{s}=510$ GeV taken in 2012 and 2013 is analyzed in this measurement. The integrated luminosities are 20 (108) pb$^{-1}$ and average polarizations are $\bar{P}_{B}=0.55\pm0.02~(0.55\pm0.02)$ and $\bar{P}_{Y}=0.57\pm0.02~(0.56\pm0.02)$ for 2012 (2013) data, where $P_{B}$ and $P_{Y}$ are the polarization of RHIC's ``Blue'' and ``Yellow''  beams. 

\subsection{PHENIX Configuration}
The PHENIX experiment consists of two mid-rapidity and two forward-rapidity spectrometers. The mid-rapidity spectrometers specialize in hadron, electron and photon identification and cover $|\eta|<0.35$ in pseudorapidity and $2\times\frac{\pi}{2}$ in azimuth.

For $\pi^{0}$ reconstruction, two high-granularity electromagnetic calorimeters (EMCal) at mid-rapidity are used. The EMCals are made up of 6 sectors of Pb-Scintillator sampling calorimeters and 2 sectors of Pb-Glass Cherenkov radiator. The calorimeters are well-suited to measure photons from $\pi^{0}$ decays. Photons are being triggered when certain energy thresholds in adjacent 4x4 blocks of EMCal towers are reached.

For luminosity measurements, beam-beam counters (BBC), two arrays of 64 quartz Cherenkov radiators with PMTs, are used which are located at $3.1<|\eta|<3.9$. To estimate systematic uncertainty from relative luminosity, a second set of luminosity detectors is needed. The zero-degree calorimeters (ZDC) which consist of W-Cu absorber and polymethyl methacrylate optical fiber Cherenkov radiator with PMTs at position of $|\eta|>6$ are used for this. Both detectors have full azimuthal coverage. While BBCs are sensitive mostly to charged particles, ZDCs predominantly measure neutral particles, in particular neutrons.

\section{Analysis Procedures}
Eq.~\ref{eq:all_def} can be re-written in terms of experimental observables as
\begin{equation}
  \ALL=\frac{1}{P_{B}P_{Y}}\frac{N_{++}-RN_{+-}}{N_{++}+RN_{+-}}; R=\frac{L_{++}}{L_{+-}}
\end{equation}
where $N_{++~(+-)}$ is the yield of $\PI$ candidates from same (opposite) helicity collisions, and $R$ is the relative luminosity of same and opposite helicity collisions.

For $R$  measurement, the luminosity miscount due to multiple collisions per bunch crossing and finite resolution of vertex width of luminosity detectors is fully corrected. As the collision rate increased especially during the 2013 running period, the effect of the multiple collisions becomes the dominant source of luminosity miscount and should be corrected. As only events with vertexes within $\pm10$ cm ($\pm30$ cm) of the nominal collision point is used in cross-section (asymmetry) measurement, luminosity miscount by the finite width resolution has been corrected.

To correct for the asymmetry and dilution due to the combinatorial background, the asymmetries are also evaluated for background events in the sideband regions below and above the $\PI$ peak (47-97 MeV/c$^{2}$ and 177-227 MeV/c$^{2}$). The actual $\ALLPI$ is then calculated from the the $\PI$ peak asymmetry ($\ALLPEAK$) and the background asymmetry ($\ALLSIDE$) as:
\begin{equation} \label{eq:back_subtract}
  \ALLPI=\frac{\ALLPEAK-r\ALLSIDE}{1-r}
\end{equation}
where r is the background fraction under the $\PI$ peak. It is obtained by Gaussian process regression. 

\section{Result and Discussion}
\begin{figure}[hbt]
  \centering
  \subfloat{
    \includegraphics[width=0.4\textwidth, height=0.2\textheight]{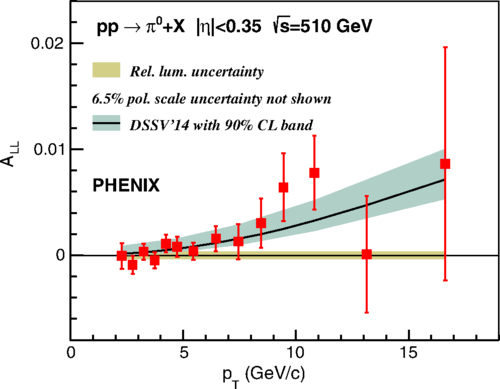}
  }
  \subfloat{
      \includegraphics[width=0.4\textwidth, height=0.2\textheight]{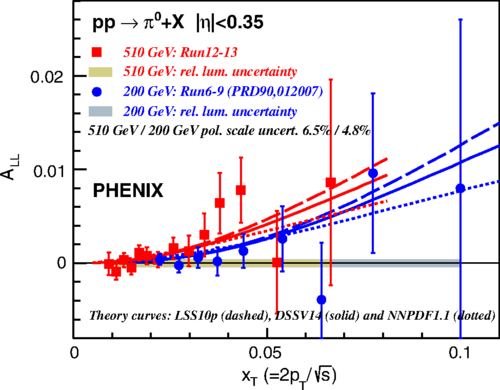}
  }
  \caption{Midrapidity $\ALLPI$ vs $\PT$ in $p+p$ collision at $\sqrt{s}=510$ GeV. Error bars are combined statistical and point-to-point systematic uncertainties. The yellow band is uncertainty from $R$. The DSSV14 theoretical curve with 90\% C.L. band \protect\cite{PhysRevLett.113.012001} is shown by green band. (left) The $\ALLPI$ vs $\XT$ (red) with $\ALLPI$ at $\sqrt{s}=200$ GeV (blue) \protect\cite{PhysRevD.90.012007} and corresponding theory curves. \protect\cite{Nocera2014276,PhysRevLett.113.012001,PhysRevD.82.114018} (right) \protect\cite{PhysRevD.93.011501}}
  \label{fig:all_results}
\end{figure}

The measurement results are shown by Figure~\ref{fig:all_results}. The world's first non-zero asymmetry in inclusive hadron production is observed. $\HGC$ in the so-far measured region via different channel at the higher $Q^{2}$ by agreeing DSSV14 theory curves which are mainly derived from STAR $\ALLJET$ at $\sqrt{s}=200$ GeV and support the positive $\HGC$. As the scale increases, the asymmtries the the same $\XT$ also increase due to DGLAP evolution. As the measurement extends the probed $x$ region, $0.01<x$, the measurement provides an additional constraint on $\HGC$ \cite{nnpdf_impact}. It is an important step to understand the spin structure of proton.

\section*{Reference}
\bibliography{Refs}

\end{document}